\documentclass [dvips, 12pt] {article} 
\usepackage {graphicx}
\usepackage {pifont}
\newcommand {\pion} {\ensuremath {{\pi^0}} }
\newcommand {\kaon} {\ensuremath {{K^0}} }

\newcommand*{\Approx}[1]{\ensuremath {\approx #1}}

\newcommand {\cd} {\normalsize \textcircled {\em \scriptsize d}}
\newcommand {\cu} {\normalsize \textcircled {\em \footnotesize u}}
\newcommand {\cs} {\normalsize \textcircled {\em \footnotesize s}}
\newcommand {\cc} {\normalsize \textcircled {\em \footnotesize c}}

\newcommand {\cdb} {\normalsize \textcircled {\em \scriptsize   $\bar{d}$}}
\newcommand {\cub} {\normalsize \textcircled {\em \footnotesize $\bar{u}$}}
\newcommand {\csb} {\normalsize \textcircled {\em \footnotesize $\bar{s}$}}
\newcommand {\ccb} {\normalsize \textcircled {\em \footnotesize $\bar{c}$}}

\newcommand {\css} {\large \textcircled {\normalsize \textcircled 
{\em \footnotesize s}}\normalsize}
\newcommand {\cssb} {\large \textcircled {\normalsize \textcircled 
{\em \footnotesize $\bar{s}$}}\normalsize}

\begin {document}
\begin {flushright}

 \emph {To my lovely Women}
\end {flushright} 
\begin {center}

{\Large Quark flowers and quark condensation \\ }
     \par\bigskip 
{\large Oleg~A.~Teplov}\par\smallskip 
Institute of metallurgy and materials science of the Russian Academy of Science, 
Moscow.
\par e-mail:\quad  teplov@ultra.imet.ac.ru \quad or \quad oleg.a.teplov@gmail.com  

\end {center}

\begin{abstract}

The mass formation of basic vector mesons up to energy 3.8 GeV 
was  studied. The investigation was done with using of jet mechanism,
the harmonic quarks and neutral colorless groups.
The stage of a whole string with a zero interquark momentum 
was considered as a separate stage of hadronization. 
The quark group with rest mass equal to experimental mass of hadron 
was named as flower. 
The quark flowers were found for $\rho$,\  $\omega$,\  $\varphi$,\  $\psi/J$,\  $\psi$(2S) and $\psi$(3770).
The flower mass of $\psi$(2S) was defined  as 3686.09 MeV.
The structure of quark flowers is symmetrical and consists 
from an aura and central part. 
The formation of quark shells, the condensation of quarks 
on the quark leaders, the floral schemes of kaon formation 
and complicated hadronization are discussed.

\end{abstract}

\section {Introduction}
\quad The present paper is devoted to study of a mass formation and a hadronization 
of the basic vector mesons in the $e^+e^-$-annihilation. In~\cite{my3, my4}
was shown, that mass spectrums of hadrons have states 
which are fully quantized by rest masses of the harmonic quarks 
and their complete oscillators. Also it is revealed that mass differences 
of hadrons are strictly quantized~\cite{my5, my7}. 

 The purpose of present article is a search of the quark groups 
 which are quantized the masses of the most symmetrical hadrons.  
 From the point of view of the author the most suitable objects 
 for examination are the symmetrical neutral vector mesons, such as, 
 for example, $\psi/J$ and other quarkoniums.

The most exact masses of the vector mesons are obtained in experiments 
on $e^+e^-$-annihilation. What happens at $e^+e^-$-annihilation?
What mechanism of formation 
     of the vector mesons from two point objects?

The modern physics answers these problems approximately as follows. 
At $e^+e^-$-annihilation, as a result of an electromagnetic interaction, 
there will be generated the intermediate virtual photon which then decays upon 
a lepton pair or a quark-antiquark pair. 
An initial quark and an antiquark should possess high momentums, but presence 
of color charges do not allow to exist it in loneliness and so the quarks fly away
together with an necessary entourage which it capture from physical vacuum. 
Thus two jets of hadrons are formed. 
Multiple creations of hadrons decrease the momentum of the leading quarks. 
Two jets "remember" also initial direction of flying away leaders.
As a minimum a jet has only one scattered hadron, for example one pion.
(We here do not view multi-jet events.)

The study bases on following experimental data and the facts:

1. Two hadronic jets practically consist of light particles 
in an energy rang of formation of the vector mesons~\cite{hoffmann}.

2. Jets have the leading quarks. 
These are the first born quarks in $e^+e^-$-annihilation \cite{hoffmann}.

3. Transverse momentums of particles in jets are small. Mean quantity of a momentum of particles \Approx 0.3 (MeV) in an energy rang of meson formation~\cite{315}.

4. In collisions of heavy nucleuses the increased birth of strangeness 
and suppression of a charm is detected~\cite{tevatron, cern}.

5. Large peaks are observed on curve of interaction cross-section at energies 
of near to masses of the vector quarkoniums.

     First four items testify that at $e^+e^-$-annihilation the heavy quarks 
     are extremely seldom formed. Even small transverse momentums testify too to it. 
Whence are occur transverse momentums? They happen from the annihilation 
a quark-antiquark pairs at formation of jets. So, the annihilation of a $s\bar{s}$-pair 
can give a transverse momentum up to 386 (MeV) to one of a hadrons, for example, to pion.
An annihilation of $b\bar{b}$-pair leads to a transverse momentums up to $\sim$5 (GeV).
However in jets a transverse momentums are restricted by $\sim$1 GeV. 
 
The curve of a interaction cross-section from energy
     of $e^+e^-$-collision has the anomalous resonant peaks in narrow ranges 
     which correspond to masses of the vector mesons, such as $\psi/J$ or $\Upsilon$(1S). 
Thus, next hadron events are observed at $e^+e^-$-annihilation 
in the low and moderate range of energies: one vector meson or two jets of hadrons. 
Certainly, in the field of the vector resonances of two jet events also are observed,
but other parallel mechanism of meson formation begins to operate.

Conserving universality of the jet mechanism, we only shall suppose 
for parallel mechanism that \textbf{bound chain} of quarks can be also generated 
in this interval; it is something like \textbf{a string of pearls or a garland}.
Viewed process essentially differs from observable monojets~\cite{monojet}
that the momentum of bound quark chain is completely compensated   and equal to zero in center-of-mass system of $e^+e^-$-pair. 
A quark chain is true single.
Such  formation of a line-up of quarks corresponds to the string model 
of quark captures from vacuum by firstly leaders 
and then by captured quarks etc~\cite{LSM}, but in this work 
we assume the string is not breaking because there goes the full conversion of 
a initial kinetic energy of leaders into quark-antiquark pairs.  
After that a whole string may be conversed into a resonant particle. 
Thus, \emph{the stage of a whole string was considered as a separate stage of hadronization}.

Taking account of all aforesaid, \textbf{ we in this paper shall search 
for neutral groups (whole string) of the harmonic light quarks 
which have total mass of rest equal to experimental mass 
of studied vector resonance}.  
As before, we shall select \emph{groups with the minimal count of quarks 
and to colorless configurations (2 or 6 quarks) with preferentially one flavor}. 
Such groups (whole strings) with a zero interquark momentum
we shall term in the further as \textbf{quark flowers} 
or is simple \textbf{flowers}.

In the present paper we shall use the same labels as in~\cite{my4}  
with the standard notation for the harmonic quarks ($d$, $u$, $s$, $c$), 
and circle-enclosed quarks of the complete harmonic oscillators (\cd, \cu, \cs).  

We begin this searching not on an empty place.
In the paper~\cite{my3} an full-color neutral 
quark configurations (which can organize temporary or stable shells in hadrons)
 were first studied with using of harmonic quarks. 
Six-color quark shells of one flavor are the most neutral of all objects 
of a QCD with respect to quantum numbers. 
From the very beginning of a birth of the harmonic quark model~\cite{my1} 
these objects were a subject of searching and examination~\cite{my3, my4, my7, my6}. 
Investigations in this area have given set of original results among
which the most important are quark configurations presented in table 1.

\begin{center}
Table 1. The earlier found quark groups ("flowers") for some mesons.
  \medskip\small 
  \begin{tabular}{|c|c|c|c|}
    \hline
 Hadron      & Quark group &The rest energy of & Hadron mass, \\
             &                    &group, MeV  &  MeV  \\
\hline
 $K^{*\pm}$  & \cs\csb\, + 3(\cu\cub) & 894.1          & 891.7   \\
   $K^{*0}$  &                        &                &896.1   \\
\hline
$a_0$(980)   & 2(\cs\csb)             & 982.66         & 984.8  \\
 $f_0$(980)  &                        &                & 980     \\
\hline
$\varphi$(1019)& 2(\cs\csb) + \cd\cdb & 1019.35         & 1019.46 \\
\hline
 $a_0$(1450)&   3(\cs\csb)            & 1474.0          & ~1474  \\
\hline
 $p\bar{p}$& 3(\cs\csb) + 3(\cu\cub)  & 1876.8           & 1876.7   \\
\hline
$D^*$(2290)$^0$& \cc\ccb\, + \cs\csb  & 2289.5            & 2290   \\
\hline
$D_1$(2420)$^0$& \cc\ccb\, + \cs\csb\, + \cu\cub & 2423.8 & 2422.2   \\
\hline
$D_2^*$(2460)$^0$& \cc\ccb+\cs\csb+\cu\cub+\cd\cdb & 2460.4 & 2458.9   \\
 $D_{s1}$(2460)$^\pm$ &                            &        & 2459.3   \\
\hline

  \end{tabular}

\end{center}

\par \bigskip

The most important results are obtained in the special work~\cite{my6} 
about full-color six-quark configuration of long-lived mesons $D^0$, $D^\pm$ and $\psi/J$. 
     Previous investigation~\cite{my3, my4}  also have shown, 
    that the number 
    of $d$-quarks in structure of particles is restricted 
    by one pair or it are not present at all (see also table 1). 
It is bound by that easiest $d$-quarks differ on properties 
from "true" $u$- and $s$-quarks \cite{my2} .
By searching groups of quark flowers we also will use one $d$-pair of quarks.

Thus, the base positions for searching were selected following:

1. Masses of $u$- and $s$-quarks in "free" and harmoniously 
bound states \cu\ and \cs \ \cite{my3};

2. Neutral six-color quark shells of one flavor (6$u$, 6\cu, 6$s$ and 6\cs);

3. Presence of the leading quarks.

The calculation of meson masses is fulfilled exclusively from positions 
of the harmonic quark model.

\section {Quark flowers}

In the given paper we shall study the vector mesons of $e^+e^-$-annihilation
approximately up to energy of $D\bar{D}$ threshold:
\begin{center}
$\rho$,\  $\omega$,\  $\varphi$,\  $\psi/J$,\  $\psi$(2S) and $\psi$(3770).
\end{center}

\subsection{\textbf{Rho ($M_\rho$=775.49 $\pm$0.34 MeV)}}

\textbf{Rho} meson is difficult object for research.
It has too greater width of decay and therefore its mass is defined 
with a significant uncertainty.  
For last years the mass of \textbf{rho} varied in tables PDG repeatedly from
770 $\pm$0.8  in 1998~\cite{Caso} up to 775.4 $\pm$0.35 MeV  
to present time~\cite{PDG}. This variation reflects the difficulties 
in interpretations of experimental data and also better understanding 
of natural processes and a improving of calculated technique. 
However data of $e^+e^-$-annihilation remained stable all this time on level 775 MeV. 
It is the author's guess~\cite{my3} the first vector resonance has six-quark $u$-shell 
with energy 633 MeV. The residual energy (142 MeV) should concentrate 
on the leaders of intact jet. This value was considered 
in~\cite{my5, CLEO} and there it correspond to a quark configuration \cd$u^0$\cdb\  
with energy 142.12 (MeV).
Preliminary flower configuration of \textbf{rho} is next:
\begin{center}
	 6$u$ + \cd$u^0$\cdb
\end{center}

\subsection{\textbf{Omega ($M_\omega$=782.65 $\pm$0.12 MeV)}}

Only one of two six-quark shells is possible in flower of $\omega$:
6\cu\  or 6$u$ with rest masses 402.75 and 632.64 (MeV) accordingly.
Hence we need to view a quark structure of the residual energies: 379.90 and 150.00 MeV.
Last value can not be represented in masses of $d$ and $u$ quarks with necessary accuracy.
On the contrary the energy 379.90 is easily interpreted by exact group \cu\cs\cub\  with a rest mass 379.92 MeV.
Then the quark flower of $\omega$ with mass 782.67 MeV has the next configuration:
\begin{center}
6\cu\  + \cu\cs\cub.
\end{center}
At this notation there is only one harmoniously bound strange quark,
but in other configuration with the same energy~\cite{my4} the $\omega$-flower will look balanced fully:
\begin{center}
	 6\cu\  + \css$\cu^0$\cssb.
\end{center}

In last notice the leading strange quarks \css\cssb\  are visible explicitly. 
Obviously the flower has the close strangeness. 
Probable leaders of the $\omega$-flower are twice harmoniously bound $s$-quarks.
Uniqueness of a $\omega$-flower is shown with calculations.
So computer search has shown, what even at inclusion in the analysis
up to 12 $d$-quarks the nearest other variant 
with the close strangeness has energy 783.48 MeV. 
This is less $\omega$ mass ~ on 7 experimental uncertainty. 

\subsection{\textbf{Phi ($M_\varphi$=1019.455 $\pm$0.02 MeV)}} 

Flower configuration of $\varphi$ is discovered earlier~\cite{my4} and given in table 1. 
It can be presented in the form of one quark chain:
\begin{center}
	 $s\bar{u}$\cd\cdb$u\bar{s}$ \quad or \quad \cd$\bar{u}s\bar{s}u$\cdb, etc.
\end{center}

What is space configuration of $\varphi$? And what are quarks the leaders? 
These problems open.

\subsection{\textbf{$\psi/J$ ($M_{\psi/J}$=3096.916 $\pm$0.011 MeV)}}

The $\psi/J$ flower in the minimum variant can contain only two six-color shells: \quad 6$s$ + 6$u$.

The difference of $\psi/J$ mass and groups is 148.923 MeV. 
It permits only two groups of leading quarks:
\begin{center}
\cu\ + \cub\ \quad and \quad \cd$u^0$\cdb
\end{center}

In the first case the residual energy will be equal 14.67 and in second - 6.80 MeV.
Thus, if not excess of energy, the meson $\psi/J$ could be interpreted 
by one of beautiful symmetrical groups:
\begin{center}
 \cu\ + 6$us$ + \cub\  \quad or \quad  \cd + 6$u$ + $u^0$ + 6$s$ + \cdb
\end{center}

However we have last and single hope to salvage a beautiful quark flower. 
The quark structure of a flower has six $us$-groups.   
The rest mass of group has energy 491.332 MeV, 
and we can replace it with real meson $K^\pm$ with the same quarks $us$. 
Then we obtain only one configuration of a $\psi/J$ flower:
\begin{center}
\cu\ + 6$K^\pm$ + \cub
\end{center}
Energy of a flower is 3096.313 MeV, but excess of energy is decreased up to 0.603 MeV. 
(The second configuration already exceeds mass $\psi/J$ on $\sim$7 MeV.) 
Remaining sequential, we shall substitute in flower the leaders \cu\cub\
on real meson $\pi^0$ with same quarks. 
As a result we obtain a beautiful flower of $\psi/J$:
\begin{center}
$\pi^0$ + 6$K^\pm$.
\end{center}

Energy of flower (3097.04 $\pm$ 0.10 MeV) is in consent with $\psi/J$ mass 
. 
Obvious leaders of the $\psi/J$ flower are harmoniously bound $u$-quarks.

\subsection{\textbf{$\psi$(2S) ($M_{\psi(2S)}$=3686.09 $\pm$0.04 MeV)}}

The $\psi$(2S) flower is discovered by analogy with $\psi/J$.
The mass of $\psi$(2S) in the minimum variant can contain only three full-color shells:
\begin{center}
6$u$ + 6$s$ + 6$u$.
\end{center}
Residual energy for leaders will be equal: 

3686.093 - 12*\textbf{105.441} - 6*385.892 = \textbf{105.448} MeV.

How funny! \emph{It differs from $u$-quark mass only in the sixth decimal sign} 
and we discover the beautiful cleanly quark flower for $\psi$(2S) 
with mass equal to 3686.09 $\pm$ 0.18 MeV:
\begin{center}
$u^0$ + 6$u$ + 6$s$ + 6$u$.
\end{center}

The neutral $u^0$ quark already figured in work~\cite{my4}. 
Here we can view it as $u\bar{u}$-pair with strictly half annihilation. 
In other words, $u$/2 + $\bar{u}$/2 are leaders.

\subsection{\textbf{$\psi$(3770)  ($M_{\psi(3770)}$=3772.92 $\pm$0.35 MeV)}}
 
If shells of $\psi$(3770) flower are the same as on $\psi$(2S) flower then residual energy 
of leaders must be equal to 192.28 MeV. 
This energy is easily interpreted by neutral symmetrical group 
of quarks with rest mass 191.87 MeV: 

\cu\cub\ + $d\bar{d}$

Hence, flower of $\psi$(3770) probably consist from following shells and quarks:
\begin{center}
$d$\cub\  + 6$u$ + 6$s$ + 6$u$ + \cu$\bar{d}$
\end{center}
It can be presented and without $d$-quarks:
\begin{center}
$u\bar{u}$ + 6\cu\ + 6$u$ + 6$s$ + $u\bar{u}$
\end{center}
The mass of $\psi$(3770) flower is 3772.51 $\pm$ 0.19 MeV and it well agrees with PDG fit~\cite{PDG}.

\subsection{Auras and center parts of quark flowers}

We should select in structures of quark flowers the central parts
 and  flower auras. 
The flower auras in mesons of $\rho$,\  $\omega$,\  $\psi/J$,\ $\psi$(2S)  and $\psi$(3770) 
are quite obvious. It is six-color homogeneous (one flavor) shells accordingly:
\begin{center}
6$u$;\quad 6\cu\ ;\quad 6$K^\pm$; \quad 6$u$+6$s$+6$u$; \quad 6$u$+6$s$+6$u$.
\end{center}
The flower auras of these mesons possesses electrical neutrality and complete color symmetry.
The central quark groups of flowers should be also neutral and symmetrical, and therefore we should write its in the next view:
\begin{center}
\cd$u^0$\cdb; \quad \css$\cu^0$\cssb; \quad $\pi^0$; \quad $u^0$; \quad $d$\cub\cu$\bar{d}$.
\end{center}
Possibly, the $\varphi$-flower $s\bar{u}$\cd\cdb$u\bar{s}$ is a quark chain and its 
also necessary to add in list of floral centers.
As opposed to inconvertible enough axial meson $\varphi$ there exist the flower states, which consist exclusively from flower auras.  
It is meson $a$0 (see tab. 1) with shell 6\cs\,  
 the resonance of $p\bar{p}$ with two shells 6\cs\ + 6\cu\ etc. 
Aura states have lesser lifetime.

In table 2 the basic results of study are presented.
\begin{center}
Table 2. The quark-flower structures of some vector mesons and calculated flower masses.  

  \medskip\small 
  \begin{tabular}{|c|c|c|c|c|c|}
    \hline
Vector       & Meson mass, & Flower mass, & Aura of      & Central part of    & Number of quarks\\
meson        &  MeV        &  MeV         & quark flower &quark flower& in aura and center  \\
 
    \hline
   $\rho^0$  & 775.49      & 774.77       & 6$u$    & \cd$u^0$\cdb &6+3 \\
\hline
$\omega$     & 782.65      & 782.67       & 6\cu    & \css$\cu^0$\cssb&6+3 \\
\hline
   $\phi$    & 1019.455    & 1019.35      &  -      & $s\bar{u}$\cd\cdb$u\bar{s}$ &6 \\
\hline
   $\psi/J$  & 3096.916    & 3097.04      & 6$K^\pm$ & $\pi^0$ & 12+2\\
\hline
  $\psi(2S)$ & 3686.09     & 3686.09      & 6$u$+6$s$+6$u$ & $u^0$ &18+1 \\
\hline
  $\psi(3770)$ & 3772.92     & 3772.51      & 6$u$+6$s$+6$u$ & $d$\cub\cu$\bar{d}$ &18+4\\ 
\hline

  \end{tabular}

\end{center}

\par \bigskip

The flower schemes of vector mesons are given on fig.\ref{fig:flo} 
{
\par
\begin {figure} [h!t!]
\begin {center}
\includegraphics [scale =0.80] {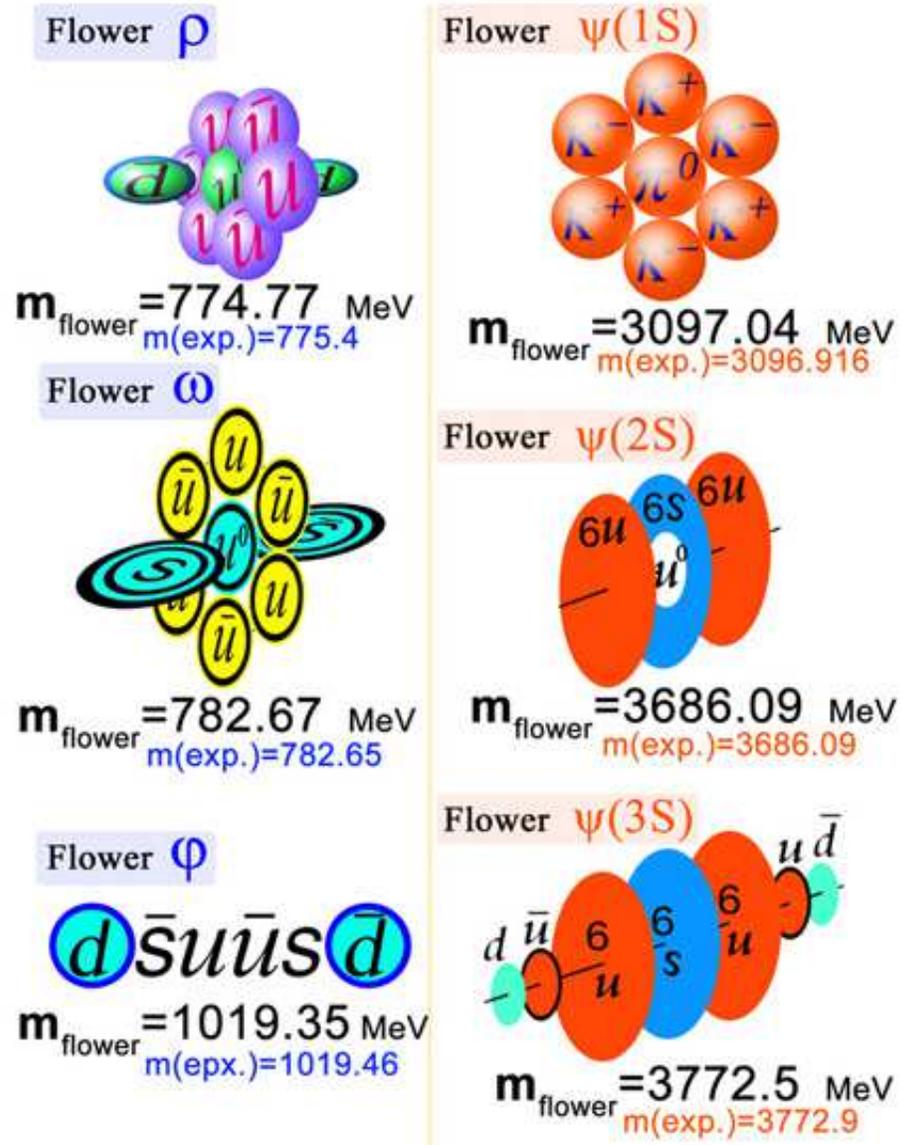}
\par
\caption [] {The possible schematic images of quark flowers.
\label {fig:flo}}
\end {center}
\end {figure}
\par

}

\clearpage

\section {Preliminary discussing of results}

\quad Discussing of unusual results  
we shall begin with a place of flower structures in the scheme of  $e^+e^-$-annihilation.
A detection of quark flowers with masses equal to masses of the vector mesons 
radically changes the scheme of its formation. 
At creation of flower the leading quarks will capture sequentially other quarks 
of a flower and its momentums will be decreased accordingly up to zero. 
 It is a transformation of the full momentum of leaders into quark masses 
 with zero momentums. 
 \textbf{Actually we deal with condensation of quarks on quark leaders}.

Also formation of quark shells on leaders at its deceleration up to zero 
can be understood as original \textbf{slowing-down radiation}.
It is completely colorless and electrically neutral. 
It is or six-quark shells, or a one-flavor quark-antiquark pairs.
 Moreover, in case of a $\psi/J$ flower there goes a condensation of kaons 
 on a pion with further transformation of this energy 
 into two heavy $c\bar{c}$-quarks and its environment.
Also similar processes should take place at creation 
of $\psi$(2S), $\psi$(3770) and probably $\omega$(783).

The simple scheme of formation of the vector mesons 
in the $e^+e^-$-annihilation now has more stages:

 $e^+e^-$ $\rightarrow$ virtual photon $\rightarrow$ leading quarks $\rightarrow$
 quark flower $\rightarrow$ quark transformation $\rightarrow$ vector meson.

It is possible to doubt in quantity and content of stages. 
For example, \emph{ we can change a stage of photon decay a  onto the leading quarks by decay onto an aura or whole flower}.
But, however, exact fit of flower and observational masses already practically does not allow to doubt in necessity of  change of prior scheme.

\subsection{\textbf{Rho}}
It is the first vector resonance with large width of decay. 
The beginning of $\rho$-peak is approximately equal to aura of a $\rho$-flower (633 MeV).
The quark structure of flower and the quantum numbers of triplet do not guess 
of a condensation of energy into a strange pair of quarks. Probably, 
after of a flower phase there goes at once a collapse of aura and formation 
of two pions from central quarks. It explains a small lifetime of $\rho$.

\subsection{\textbf{Omega}}

The width of $\omega$-peak is equal to $\sim$8 MeV. The beginning of peak is $\sim$774 MeV 
that above a threshold of formation of ss-pair on $\sim$2 MeV.
The structure of a $\omega$-flower has a twice coupled strange pair of quarks 
and it is possible to guess, that after of a flower phase the condensation 
of energy can occur into a pair of the strange quarks in the vector variant.
It explains an origin of the singlet state and significant lifetime of $\omega$. Perhaps also \textbf{just the stage of condensation leads  to anomalous peak increase of cross-section}. 

\subsection{\textbf{Phi}}

 Possibly $\varphi$ is a quark chain ("flower without aura")  
and its quark structure can be written down
in two symmetrical variants:
\begin{center}
\cd\csb\cs\csb\cs\cdb\ \quad or \quad \cd$\bar{u}s\bar{s}u$\cdb.
\end{center}
Not clearly, what is the leaders of chain? Presence $s\bar{s}$ quarks
in a flower provides to meson $\varphi$ a closed strangeness 
and a quark condensation is not required.
However the annihilation of \cd\cdb-pair with redistribution of energy 
in the kinetic and color fields is possible.
So far question about the quark structure of $\varphi$ remains open.

\subsection{\textbf{$\psi/J$}}

Uncertainty of $\psi/J$ mass is less 9 time than for a flower. 
If flower concept is true the mass $K^\pm$ can be calculated more exactly 
with using the masses of $\psi/J$ and $\pi^0$:
\begin{center}
 $M_{K^\pm}$ = 493.656 $\pm$ 0.002 MeV.
\end{center}
The calculated value agrees with experimental data, 
but better with data \emph{no Denisov}~\cite{noDEN}.
  
The kaon condensation on pion requires a careful analysis. 
A formation of two heavy quarks due to condensation of kaons 
on $u$-oscillator of pion arouses the certain doubt. 
The matter is that though the mass of charged kaon consists 
almost of masses $u$- and $s$-quarks, nevertheless, there are 2.34 (MeV)
which relates to kinetic and coupling energies of quarks. 
As shown in~\cite{my2} the velocity of $u$-quark 
in a kaon achieves 0.2 from velocity of the light. 
It is difficult to present, how condensation can be executed  
at high velocities of quarks? 
Certainly, the relative velocity of $u$- and $s$-quarks can be $\sim$0 at some moments.

Nevertheless, this problem was leading to idea about the possible 
flower solution for kaon pairs $K^+K^-$.
Searching has led to following result:

\begin{equation}\label{Kaplus}
M_{K^+K^-} = 2 * (6M_u) - 2M_u - M_{\cu^0}=2*493.643 (MeV).
\end{equation}

Importance of the equation~(\ref{Kaplus}) is obvious. 
The mass of kaon pair can be also effect of condensation 
of two 6$u$-shells except the annihilating group $u\cu^0\bar{u}$.
The scheme of $K^+K^-$ creation by condensation of $u$-quarks is given on fig.2. 
{

\par
\begin {figure} [h!t!]
\begin {center}
\includegraphics [scale =0.50] {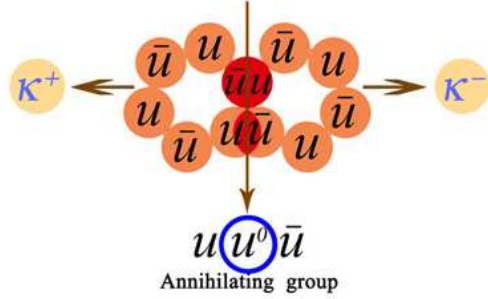}
\par
\caption [] {The flower scheme of $K^+K^-$ creation.
\label {fig:kk}}
\end {center}
\end {figure}
\par
}

The $\psi/J$ flower and also equation~(\ref{Kaplus}) open the way 
for an operations with kaon condensate. 
Actually under kaon condensation there  can be hidden the $u$-quark condensation.
 The calculated value for mass of $K^\pm$ (493.643) is also in a fine consent
 with experimental data \emph{no DENISOV}~\cite{noDEN} which equal to 493.642 MeV. 

\subsection{\textbf{$\psi$(2S)}}

The $\psi$(2S)  is formed of cleanly quark flower (see fig. 1). 
It differs from other flowers by the minimal central part $u^0$. 
Neutral centres on the basis of a $u$-quark have also the $\rho$ and $\omega$.
About $u^0$ we spoke earlier in~\cite{my4}. 
The neutral $u^0$ can be represented as $u$/2 + $\bar{u}$/2, i.e.
by $u\bar{u}$-pair with 50\% an annihilation.
However it is more probably the center and aura are forming together or the aura even first. 
After that quarks of aura condense onto the $u$/2 + $\bar{u}$/2. 
If all the same to operate concept the leading quarks then $u^0$ can be considered 
as two leading quarks, for example $u\bar{u}$, in the virtual sense.

The mass error of $\psi$(2S) is 0.04 ($\sim$0.001\%)
that is 5 times less, than error for flower mass.  
The masses of harmonic quarks are recalculated with using mass 
of $\psi$(2S) and are given in the table 3.
Though values of quark masses have remained the same, 
but mass errors of quarks considerably decrease 
if the floral concept is perfectly correctness. 

{
\begin{center}
Table 3. The quark masses from $\psi$(2S) mass and its flower.  

  \medskip\small 
  \begin{tabular}{|c|c|c|}
    \hline
Quark  &  Quark mass~\cite{my3}, MeV & Quark mass (this work), MeV \\ 
\hline
$d$    & 28.811  $\pm$ 0.0014    & 28.8107  $\pm$ 0.0003   \\
\hline
$u$    & 105.441 $\pm$ 0.005     & 105.4412 $\pm$ 0.0010   \\
\hline
$s$    & 385.89  $\pm$ 0.019     & 385.8929 $\pm$ 0.0036   \\
\hline
$c$    & 1412.30 $\pm$  0.07     & 1412.288 $\pm$ 0.013    \\
\hline
$b$    & 5168.7  $\pm$  0.26     & 5168.68  $\pm$ 0.05     \\
\hline

  \end{tabular}

\end{center}
}
\par \bigskip

\subsection{\textbf{$\psi$(3770)}}

The flower of $\psi$(3770) has the aura of $\psi$(2S), but clearly expressed axial quarks.
 
\subsection{\textbf{6$u$-flower}}

\quad Earlier in~\cite{my3} exclusively important energy equality has been found,
the rest mass of 6$u$-shell is equal to the mass sum of the \pion and \kaon.
 It means that \kaon mass ($M_{K^0}$(exp.)= 497.648 $\pm$ 0.022 MeV~\cite{PDG}) can be expressed 
 also by the floral concept:
\begin{equation}\label{Ka0}
M_{K^0} = 6M_u - M_{\pi^0} = 497.671 \pm 0.003(MeV) 
\end{equation}

As the 6$u$-shell is completely neutral and also a pion is truly neutral meson
this equality~(\ref{Ka0}) informs about a potential neutrality of kaon.
In fact complete neutrality kaon is expressed in oscillations $K^0$ and $\bar{K}^0$.
This process can be noted on flower mode:

\quad \quad \quad $K^0$ $\longleftrightarrow$ $K^0$-flower $\longleftrightarrow$ $\bar{K}^0$.

The energy equal to mass of a kaon cannot exist long in the form 
of the virtual flower ($M_{6u}$ =632.645 MeV) because of the uncertainty relation 
($\Delta$E = $M_{\pi^0}$*c*c). This energy should be immediately conversed to quarks:

\quad or \quad $s$-quark + etc., \quad or \quad $\bar{s}$-quark + etc.

[ Certainly quantum transition through an energy barrier 632.64 MeV 
may be studied in respect of the $K$-oscillations, but we need remember 
that there can be a transition through \cs\csb-oscllator directly~\cite{my3}.] 

We could rewrite equality~(\ref{Ka0}) onto a mass of pion, but because of $K$-oscillation the similar equation would be only formal 
and would contradict the status of a pion, as to truly neutral particle 
with the least hadronic mass. 
Just least mass among of hadrons  is the cause of the special status \pion --
truly neutral \cu\cub-oscillator.
Other quark flavors have two neutral mesons in the ground states 
($K^0\bar{K}^0$, $D^0\bar{D}^0$, $b^0\bar{b}^0$). 
On the contrary the equality~(\ref{Ka0}) has deep physical sense. 

\section {\textbf{Deductions}}

1. The bound quark chains are truely one-jet events.

2. The quark flower is the bound quark chain with zero interquark 
momentum and the rest mass equal to the mass of vector meson.

3. Here was opened the quark flowers of next vector mesons: 
$\rho$,\  $\omega$,\  $\varphi$,\  $\psi/J$,\  $\psi$(2S) and $\psi$(3770).

4. Here was discovered the quark condensation.

5. The heavy quarks can be formed by condensation of light quarks.

6. The process of condensation can be by cause of anomalous peak increase of interaction cross-section.

\section{\textbf{Conclusion}}

\quad Obviously, the investigations with using of the new tool
(harmonic quarks) has led us to the active intrusion 
into a complex field of a hadronization,
a field which is practically closed for exact studies. 
This field is not accessible to modern methods because 
these methods have not of other such  exact instruments 
as harmonic quarks.

In following article the results of floral study will be represented 
for next hadrons:
 
$\psi$(4040), $\psi$(4415), $D^{*0}$, $D_s^*$, $\Sigma_c$, $\Upsilon$(1S), $\Upsilon$(2S) etc.

\end {document}